# Analyser Framework to verify Software Components

**Rolf Andreas Rasenack**

FH Oldenburg/Ostfriesland/Wilhelmshaven        De Montfort University,
Fachbereich Technik, INK        Software Technology Research Laboratory
Constantiaplatz 4        The Gateway
26723 Emden        Leicester LE1 9BH, UK
rasenack@technik-emden.de

**Abstract:** Today, it is important for software companies to build software systems in a short time-interval, to reduce costs and to have a good market position. Therefore well organized and systematic development approaches are required. Reusing software components, which are well tested, can be a good solution to develop software applications in effective manner. The reuse of software components is less expensive and less time consuming than a development from scratch. But it is dangerous to think that software components can be match together without any problems. Software components itself are well tested, of course, but even if they composed together problems occur. Most problems are based on interaction respectively communication. Avoiding such errors a framework has to be developed for analysing software components. That framework determines the compatibility of corresponding software components. The promising approach discussed here, presents a novel technique for analysing software components by applying an Abstract Syntax Language Tree (ASLT). A supportive environment will be designed that checks the compatibility of black-box software components. This article is concerned to the question how can be coupled software components verified by using an analyzer framework and determines the usage of the ASLT. Black-box Software Components and Abstract Syntax Language Tree are the basis for developing the proposed framework and are discussed here to provide the background knowledge. The practical implementation of this framework is discussed and shows the result by using a test environment.

## 1 Introduction and Motivation

Component-based software technology represents a software production paradigm that concentrates on the reuse of software components to develop large software systems. The reuse of software components, even so called components-of-the-shelf (COTS), to assemble applications are in practice often problematic. It was hoped that software components can be match together without any change [SH04]. But often in practice the behaviour of a software component is not the same as expected. Due to incompatible interfaces for communication/ interaction between software components and the lack of functionality this problem occurs.

The circumstances that software components cannot be reused ''as-is'' is identified by many researchers. Therefore software components have to be analysed whether they can be match, be adapted or short it is necessary to verify their compatibility. With the assistance of an analyser framework for software components such problems will be visible and an appropriate reaction can be performed. That framework determines the compatibility of corresponding software components and can be used as a part of the adaptation framework described in [Ras08]. A promising approach to develop an analysing framework for software components is applying an abstract syntax language tree (ASLT) [Wol07, W+04, Y+04]. The ASLT is the hierarchical representation of object-oriented structures and provides the appropriate information. With their assistance associations and couplings between software components can be compared and proofed.

## 2 The Nature of Black-Box Software Components

At first black-box software components nature has to be discussed since the analysing process is based on that software components. Chapter 2.1 describes the definition of black-box software components. The following discussion clarifies that classes in an object-oriented programming language, like Java, can be seen as software components on condition that classes are logically coherent. The following chapter 2.2 defines the three elements (*component, component interface, component specification*) of software components. Chapter 2.3 discusses addressed problems if software components will be connected.

Software components have some properties and can be characterized by a definition. The term software component is defined in literature in manifold ways. Some definitions try to define the term software component in a general way without technical considerations. Other definitions concentrate on the context in which the software components can be used. For instance software components can be seen as parts of a software system or they can be seen as service provider. To cover all aspects that are related to software components in different context is probably not exhaustive possible. Therefore we concentrates on the most convinced definitions in this topic and excerpt a definition for adaptable software components.

## 2.1 Definitions

As one outcome of the first Workshop on Component-Oriented Programming 1996 (WCOP´96) at European Conference on Object-Oriented Programming 1996 (ECOOP´96) in Linz, Szyperski and Pfister developed the following definition of the term software component:

"*A software component is a unit of composition with contractually specified interfaces and explicit context dependencies only. A software component can be deployed independently and is subject to composition by third parties.*" [Muh97]

In other words this definition describes a software component which consists of combinable pieces software. Pieces of software for instance in the object-oriented programming language Java [***03a] can be a class. This implies that a software component is more coarse-grained than a single class. Logically coherent classes can be compounded to a software component. Well defined interfaces of software components described by a contract are a necessary premise for communicating between software components. A contract, between a developer and a client is a precise specification attached to an interface. It covers functional and extra-functional aspects. Functional aspects include the syntax and the semantics of an interface whereas the extra-functional aspects include the quality-of-service guarantees [Szy02].

Additionally software components are designed not only for domain specific applications. They encapsulate its implementation so that it is not possible to have access to the construction details and therefore software components are self-contained. Szyperski abstract this definition into a technical part with considerations such as composition, independence, and contractual interfaces and a market-related part with considerations such as deployment, and third parties [Szy02]. This reflects the practical benefit for the development process of software components.

Another important definition comes from Sametinger. In contrast to the above mentioned definition Sametinger gives a more general definition without consideration of market-related aspects. As one result, in the following definition it is stated that software components are any reusable artefacts. The used term artefact represents different forms of software components. This can be source code or a black-box view that hides the internal details of a software component for instance.

"*Reusable software components are self-contained, clearly identifiable artefacts that describe and/or perform specific functions and have clear interfaces, appropriate documentation and a defined reuse status*". [Sam97]

Self-contained software components mean, in Sametingers definition, that a software component has its own functionality and do not need additional software components or services to provide this functionality. Furthermore software components should be contained in a file and not being spread over many locations then it is identifiable. It has a clear defined interface that hides details that are not needed for reuse. The documentation (specification) must provide enough information to retrieve a software component from a repository, gives information in which context this software component can be used, make adaptations possible. Furthermore the mentioned reuse status of Sametingers definition provides release information of the software component.

The definitions discussed here, include only two representative definitions. But the term coupling between software components are not considered. The necessity to consider the notion coupling, is caused by flexible combining of software components especially for adapt them. In [WY03] the term coupling was taken into account and describes the level of dependencies between interacting software components. Coupling between software components will be differentiated into low coupling or high coupling. The design of highly-coupled software components is based of assumptions between them. Assumptions include for instance every time availability of corresponding software components, syntax for invoking the functionality of interacting software components or data exchange between the software components has to be done every time in the same format. Advantageous of this highly-coupled software components are increasing the performance between the related software components. Disadvantageous is the fact that the software components are specific designed to communicate to each other. This means if requirements are changing for instance in the direction of functionality then all related software components have to adjust to the new situation. But in sense of adaptation of software components it is not acceptable to redesign all related software components because of additional costs, time and may be putting errors in the new developed software. Therefore low coupling is a preferred approach in which software components operate extensive autonomous via interfaces and does not need to be concerned with other software components internal implementation. This is important because changes in one software component have no influence to the corresponding software component. Thus the approach of low coupling is necessary to consider in the definition of software components which can be adapted.

Derived from the above-mentioned discussion, the following combination of definitions will be considered in the area of adaptation of software components:

A *software component is a piece of software which offers a coherent functionality and exhibits certain autonomy by strict encapsulation of the implementation. Flexible combining and separation of software components are achieved by low coupling. Well defined interfaces, responsible for the communication and interaction between components, include a specification which additional describes the behaviour of the software component. The internal structure of a software component will not be considered. Software components can be composed of single software components to achieve an extended functionality.*

## 2.2   Elements of Software Components

The structure of software components characterizes different elements of software components. Yang and Ward [WY03] define five elements of a software component. That includes code, specification, interface, design and documentation. We focus on the approach with the abstract view on three structure elements of a software component. These are:

- Component
- Component interface
- Component specification.

Figure 1 shows a software component with its typical three elements and its simple model representation. A well defined *component interface* is required for communication and interaction with other software components. It separates the software component to each other and is described by a corresponding *component specification*. Incoming and outgoing information/ services of software components will be processed by the appropriate provided and required interfaces. The *component* is an element that hides its internal structure for using of third parties. That is, it provides the internal logic (e.g. classes in object-oriented programming) which is not present for the client. Hence a component represents certain behaviour and is addressed by the component interface.

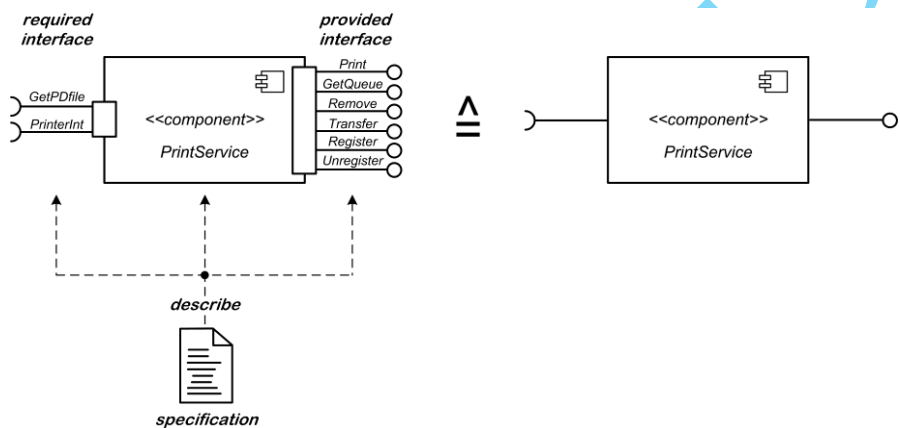

Figure 1: Elements of a Software Component [7] with its simple Model

Software components are represented in different views depending on their abstraction level. The abstraction level defines the different alternatives of the access to the structure of software components. They can be distinguished into black-box, white-box, glass-box and gray-box software components. The scope of research is directed on black-box software components. For instance a binary form of JavaBeans [***03b] can be a representation of a black-box software component.

## 2.3   Component Mismatch

The increasing productivity of the software development process is attended by the ability of reusable software components to combine (compose) them.

Composing applications out of reusable software components leads to rapidly developing in contrast to developing software from scratch. However systematic development of applications from existing software components is an elusive goal. The reason for that is caused by:

- The inability to locate the desired software components
- The lack of existing software components
- Mismatches between software components to build applications.

To solve the problem of the inability to locate the desired software component it is necessary to provide a component pool that catalogues and categorizes the software components. So that is possible to retrieve a software component for the desired needs. The lack of existing software components leads to development of appropriate new software components. It is obviously that this new components have to store into the component pool. The paper, cited by [Ras08] discusses an approach that includes a component pool.

Reasons that software components cannot interoperate are described by Shaw [Sha95]. To them belong different assumptions about how data is represented, how they are synchronized and what semantics they have.

## 3  Abstract Syntax Language Tree

Source code of a programming language typically consists of instructions stored in a text file. Additionally in object-oriented programming languages hierarchical structures are defined too. Software projects can have a certain amount of separated files. This leads to unclear programming structures and the developer lost the overview. Just in the analysis of source code it is very difficult to find irregularities and errors. The developer can have important strategic advantages by administration of source code by using an ASLT. This chapter describes the concept and the usage of the ASLT. Advantageous is that the source code file is synchronized with the model presented by the ASLT. This means no information is lost by transforming from source code to ASLT and vice versa. The ASLT for the programming language Java consists of the appropriate API and adequate tools for transforming between source code and ASLT view.

## 3.1  The ASLT Concept

The concept and the implementation of the Abstract Syntax Language Tree (ASLT) is a collaborative work [Wol07] and are designed for source code manipulation of an application. With the assistance of an ASLT the processing (analysis) of software components shall be conducted.  The ASLT is the representation of object-oriented structures (packages, classes, variables and methods), which become visible as hierarchical elements (nodes). These nodes are depicting in Figure 2. It is a graphical representation of the Java source code `TestBed.java` after transformation into `TestBed.aslt` by using the tool *CodeToASLT*. This transformation tool is part of the ASLT build. To show the hierarchical structure of Figure 2 a viewer tool is necessary. It is named ASLT viewer.

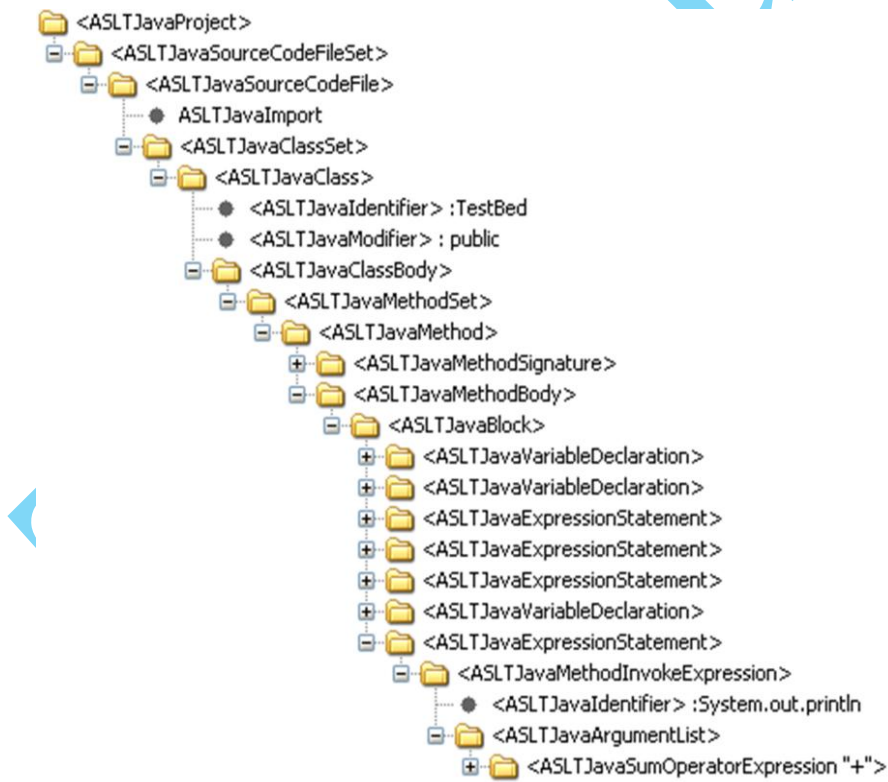

Figure 2: ASLT Tree

The ASLT is the basis for variants of implementation and/or views (UML-class Diagram, UML-Sequence Diagrams, Component-Diagrams etc.), which are made available to the developer (Figure 3). Each view offers to the developer a special sight of a project. Thereby only certain parts of a project will be represented, the remaining other parts becoming invisible by folding (compare the ASLT tree Figure 2). The ASLT is the model for the administration of hierarchic elements and particularly for the representation and/or finding of meta-information, which is intended for semantic check of software components [WYSRA04, YWSAR04]. In this article, meta-information is not the subject of discussion.

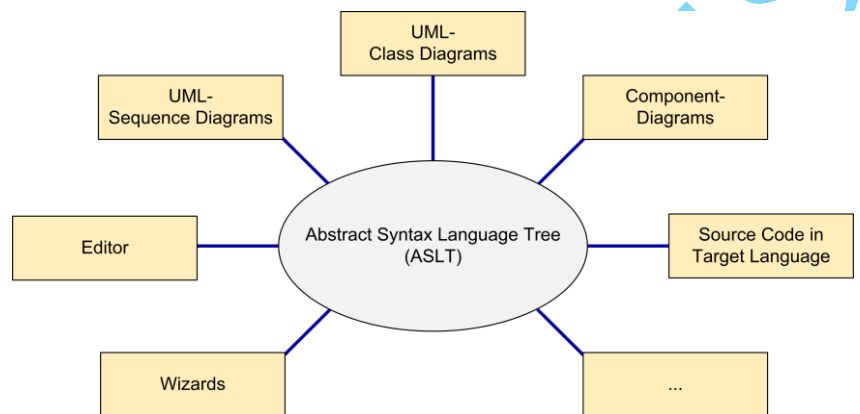

Figure 3: Use of the ASLT [WSYSR04]

## 4  Proposed Analysing Framework

The architectural design of a framework for analysing software components provides an environment in which black-box software components are checked whether they can match or not. The principle of that check is based on identifying relations and dependency between software components. Consequently the compatibility of corresponding software components will be proofed. This chapter describes in general the approach to cope with that components and show how the framework operates.

As mentioned before software components can be seen as a self-contained unit with an appropriate interface for communication to its environment. In the literature such software components are named black-box software components. Such a black-box software component has certain functionality, for instance it calculates the capacity of containers and provides the result for further processing. Advantageous of black-box software components is its reusability. Software components can be stored in a data base to have a pool of software components with that it is possible to compose large applications. That saves development time and costs.

To simplify the understanding process of the analysing framework, we consider black-box software components as classes of an object-oriented programming language (Java [***03a]). Existent relations and the communications between classes, that will be applied in larger structures (e.g. applications), must be consider because exchanging classes can have different communication structures. Within integrated development environments (IDE´s) such relations will be proofed during programming. The IDE will advise programmers on errors, for example like declaration of a wrong return type, by displaying information on the computer screen. But what happens by composing software components? The IDE has no influence during composing. To guarantee the compatibility of software components (classes) we introduce an analyser framework. Its task is to proof corresponding classes, and to react on exchange of classes or modifications. With the assistance of the analyser framework the composing process will be check and it provides error correction.

The user of the analyser does not know the internal details of the framework essentially, because the framework provides interfaces for communication and offers the result of the test process. That means the analyser framework is easy to use. Pre-condition for developing an easy to use framework for analysing classes is applying the concept of the ASLT. [Wol07, Wol06].

## 4.1 Proposal

The concept to realise the analyser framework contains the verification of communication between software components. Based on the object oriented programming language Java we concentrate on classes as representative of software components. This happens for simplifying the understanding process. The communication between classes can be seen as the access to an object of another class or the transfer of parameters.

Figure 4 depict the communication between three classes. The class SampleClassA calls two methods from two different other classes. Those are named SampleClassB and SampleClassC and provide the appropriate methods. The communication between these classes is clear and the compatibility is available.

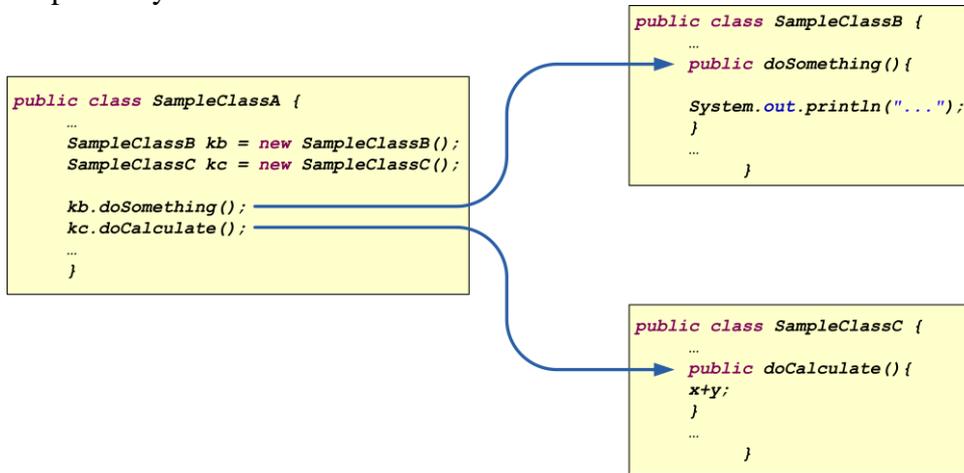

Figure 4: Communication between Classes

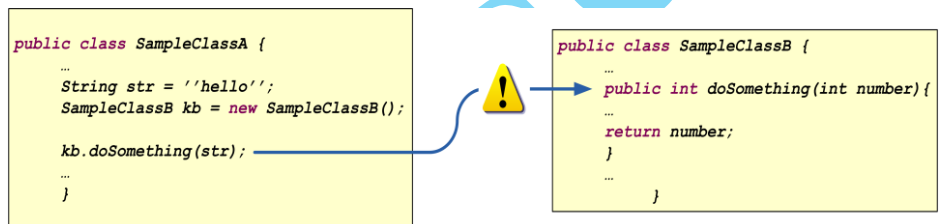

Figure 5: Sample of Communication Problem between Classes

A different sample of communication between two classes is shown in Figure 5. It is assumed that both classes working independent as software components and shall be composed together. The class SampleClassA tries to call the method doSomething(str). The parameter *str* is from type *String*. In contrast to class SampleClassA with its method doSomething(str), the method doSomething(number) of the class SampleClassB is implemented with the parameter number from type int. Obviously there is a communication problem. Problems as described here can be avoided by analysing the corresponding classes. This means all related classes have to be taken into account to find associations between software components.

The query of related classes respectively software components (Figure 6) about the relations delivers the necessary information for the analyser framework. As mentioned before the task of the analyser framework is to verify the compatibility of software components.

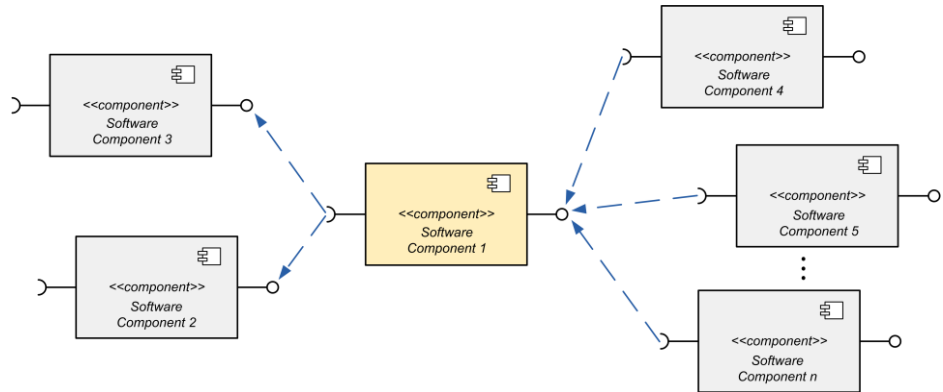

Figure 6: Relation between Software Components

Figure 4 and Figure 5 show the communication between classes based on the programming language Java. In other words software projects are organised in a hierarchical structure. Packages, classes, methods etc. are hierarchical structure elements as known in object-oriented programming environments. Classes can be seen as software components. The analyser framework will do its work after one or more software components of a project are substituted, adapted or modified. It looks on separate views of classes. A Java class can have different occurrence. The source code of a Java class is only the textual representation of the syntax of Java. However the compiled version is named as a binary representation of the Java class. Both versions will not represent hierarchical structures. But this is necessary for the analyser framework because this view on a Java class gives the information of communication between related classes and provides the possibility to manage and manipulate with that Java classes.

With the assistance of the Java Reflection Application Programming Interface (API) associations between software components will be located during run-time. This is necessary if the used class is not defined during compile time or information about that class has to collect during run time. This means the Refection API is able to collect information about classes, super classes, implemented interfaces, arrays, methods and attributes. The class under test is represented by a java.lang.Class object. The Java class Class is the main basis of Java Reflection. The analyser framework takes the collected information and stores it into a Java Vector.

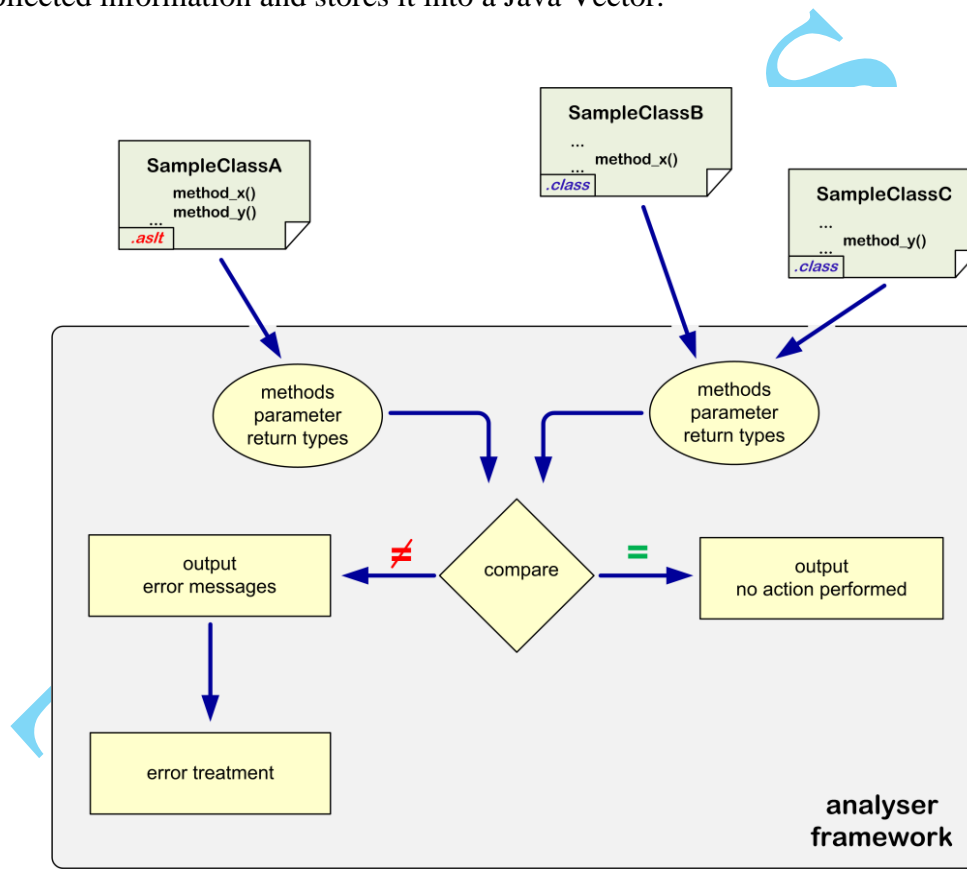

Figure 7: Analyser Framework and its Process

Additional a list of classes which communicates to the class under test is required. For that reason Java classes will be transferred into the hierarchical structure of the ASLT. After transferring into the ASLT form the information will be stored into a second Java Vector. Now information of called methods, transfer parameters and expected return types are available. This information will be compared with the information of methods, parameter and return types collected from the Reflection API. The result of comparing is stored into a result Java Vector. If inconsistencies appear during that compare process an error message will be created for further operation. Figure 7 depict the functionality of the analyser framework.

## 5    Analyser Framework Implementation

This chapter describes the practical implementation and the functionality of the analyser framework. For better understanding the appropriate classes and methods are shown in an UML diagram. This software project is separated into the framework part (Figure 8), which consider project information from an outsourced file (Figure 9) and the test environment part (Figure 10).

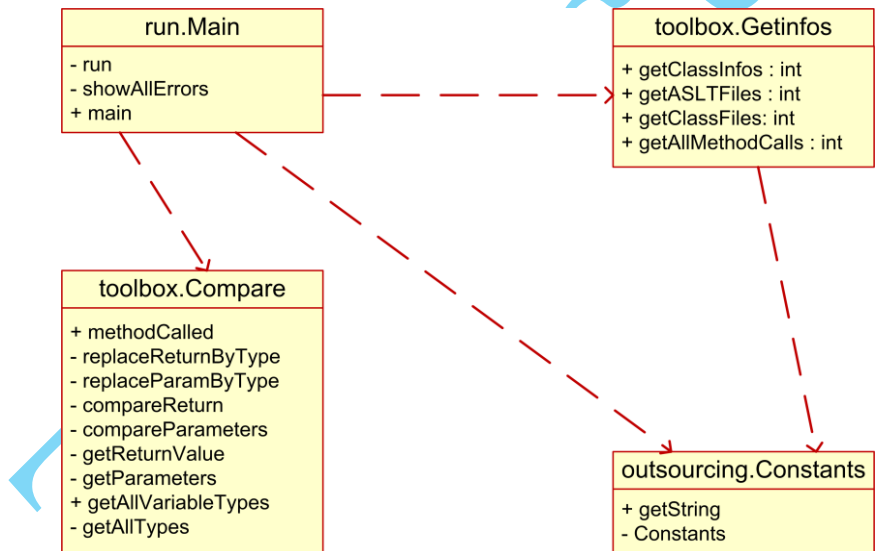

Figure 8: UML Diagram of the Analyser Framework

### 5.1  Overview Analyser Framework

This chapter gives a short overview of implemented Java classes of the analyser framework. It consists of the following four Java classes:

1. `run.Main`
2. `toolbox.GetInfos`

3. `toolbox.Compare`
4. `outsourcing.Constants`

The association between that Java classes and the implemented method are depict in Figure 8. The class `run.Main` is used as access point to the analyser framework and it contains the `main` method. The path to the software project will be read and the Java Vectors for storing search information will be implemented. The class `toolbox.GetInfos` is responsible for collecting information about classes within that software project by using the Java Reflection API and the ASLT API (see Figure 7). The search provides the information about methods, parameter and return types and will be compared in the class `toolbox.Compare`. The class `outsourcing.Constants` represents an interface to the file `constants.properties` (see Figure 9). That file defines properties to configure the project.

```
PathToApplication=E:/Eclipse/Analyser/DUT/test1
ASLTFileExtension=.aslt
ClassFileExtension=.class
ASLTJavaExpressionStatement=ASLTJavaExpressionStatement
ASLTJavaIdentifierExpression=ASLTJavaIdentifierExpression
ASLTJavaLiteralTag=ASLTJavaLiteralTag
ASLTJavaMethodInvokeExpression=ASLTJavaMethodInvokeExpression
ASLTJavaSimpleAssignmentOperatorExpression=ASLTJavaSimpleAssignmentOperatorExpression
ASLTJavaVariableDeclarator=ASLTJavaVariableDeclarator
ASLTJavaVariableDeclaration=ASLTJavaVariableDeclaration
DebugLevel=1
```

Figure 9: Properties of Configuration File `constants.properties`

The functionality of the analyser framework will be proofed by a test environment that is depicting in Figure 10. It consists of three Java classes, TestBed, SampleClassA, and SampleClassB. The method main() to starting the test environment is implemented in the Java class TestBed. SampleClassA and SampleClassB provide several methods to communicate each other.

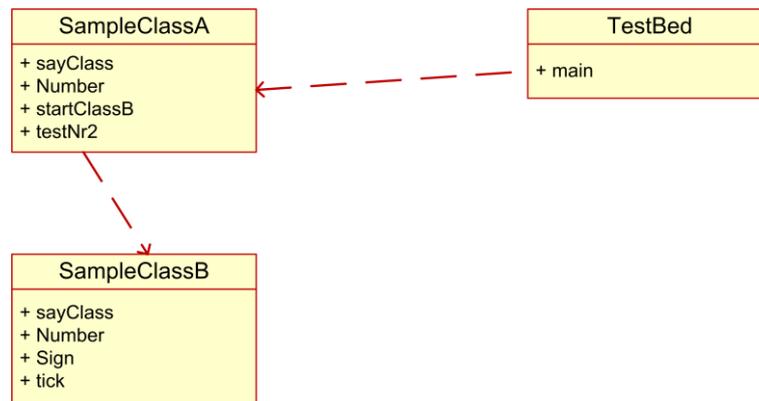

Figure 10: UML Diagram of the Test Environment

Now the developer is able to manipulate the parameters or return types of methods etc. to simulate an error. This error will be recognised by the analyser framework and presents an error message on the computer screen.

## 5.2  Java Class Description

Before applying the analyser framework it is necessary to make some adjustments of project properties. This will be done in the file constants.properties. The path to the test environment is configured by assistance of the property PathToApplication. The property DebugLevel defines the output level of search results and is represented by a number:

0: debug level is deactivated, no output information is given
1: essential class information (methods, parameter, and return types) will be created and are available as output information on the computer screen
2: all .aslt and .class information will be shown on the computer screen

Different other properties define the type of file which has to be considered and define the names of ASLT nodes which gives the information for searching methods, parameter, and return types.

The analyser framework will be running by starting the main()method in the Java class run.Main. The DebugLevel and the PathToApplication properties will be read at first. The names of the project files with the extension .aslt and .class are stored into the appropriate Java Vector. The method getClassInfos() contained in the Java class toolbox.Getinfos will be called. With the assistance of the Java Reflection API gives that method information about methods, parameter, and return types for storing into a Java Vector. That necessary procedure is used to compare information that is collected from the ASLT tree. For this purpose the method getAllMethodCalls() collects the information. The search within the ASLT tree gives the appropriate information of methods, parameter, and return types. The ASLT tree represents the hierarchical structure of a software project therefore it shows much information. To decrease the amount of search information it is recommended to specify an access entry by using the property ASLTJavaMethodInvokeExpression. Below this node within the ASLT tree method calls are present. The ASLT API provides a method getMethod()that gives the information of method calls within the ASLT tree in the following form:

<ASLTJavaMethod>: class.method name

The resulting information about that search will be stored in a Java Vector. The analyser framework calls the method getAllVariablesTypes() that is implemented in the Java class toolbox.Compare. Together with the method getAllTypes() a list with all variables from each ASLT-file (file with extension .aslt) will be generated. To obtain the types of the variables the ASLT-file will be scanned to find the entry ASLTJavaVariableDeclaration that is defined as property in the configuration file constants.properties. The result of this search provides the desired list. The collected information will be compared by using the method methodCalled() implemented in the Java class toolbox.Compare. All method calls of the ASLT-files will be compared with the appropriate .class-files. The analyser framework will compare the parameters and the return types of the .aslt-files and .class-files in the case of conforming method calls. Otherwise failures that occur during the comparing process will be stored in the Java Vector allErrors. The analyser framework stores the following information in the Java Vector:

ASLT class names
called .class-file
called method
expected and given parameter
expected and given return types

The method showAllErrors()implemented in the Java class run.Main
returns all error messages that occurred during the comparing process on the
computer screen. This last operation of the anlyser framework provides the
result to software developer. Due to that result the developer is able to
compose verified software components.

**Conclusion**

The use of reusable software components is advantageous in sense of
effective programming applications. Furthermore costs of software
development are calculable and therefore an extreme favourable alternative
to developing applications from scratch. Problematic of using software
components is their behaviour during composing.

Reusable software components will be used due to the use of new
technologies, error correction (e.g. mismatched interfaces) and
implementation of newer functionalities for example the fulfilment of user
requirements. In most of applications necessarily one or more software
components of an application has to be adapted or software components
must be added. But at least they have to analyse to verify their
compatibility. The described framework assists the developers work by
analysing software components. The application is submitted to an
analysing process and proof the coupling of related software components
and their functionality. This will be done by a framework. With the
assistance of this framework the goal is pursued of proving compatibility
and on a long-term basis to provide the functionality of the application.
Furthermore the aim is to be carried out a contribution to program software
products reliable in service.

The software development process used with the here presented
concept will be more transparent because of a comparison algorithm that
makes sure that software components can be match together. The
comparison algorithm uses information of Java classes that are transformed
into the ASLT. Without the assistance of the ASLT it is very extensive and time-consuming to manage the analysing process. Favourable at this proposed framework is reducing development time and avoiding inconsistencies between software components.